\pdfoutput=1
\documentclass[11pt]{article}
\usepackage[T1]{fontenc}
\usepackage[utf8]{inputenc}
\usepackage{lmodern}
\usepackage[margin=0.82in]{geometry}
\usepackage{amsmath,amssymb,bm}
\usepackage{graphicx}
\usepackage{booktabs,tabularx,array}
\usepackage{microtype}
\usepackage{xurl}
\usepackage[hidelinks]{hyperref}
\usepackage{placeins}
\usepackage{float}
\usepackage{enumitem}
\newcolumntype{Y}{>{\raggedright\arraybackslash}X}
\newcolumntype{C}{>{\centering\arraybackslash}X}
\newcolumntype{R}{>{\raggedleft\arraybackslash}X}
\title{Quantum-Classical Fragmentation with the Effective Fragment Molecular Orbital Method}
\author{Federico Zahariev$^{1,2}$, Vassiliki-Alexandra Glezakou$^{3}$, and Mark S. Gordon$^{1,2}$}
\date{}
\begin{document}
\maketitle
\begin{center}
\small $^{1}$Department of Chemistry, Iowa State University, Ames, Iowa 50011, USA\\$^{2}$Ames National Laboratory, Ames, Iowa 50011, USA\\$^{3}$Chemical Sciences Division, Oak Ridge National Laboratory, Oak Ridge, Tennessee 37831, USA
\end{center}
\vspace{0.5em}
\begin{abstract}
We present the quantum effective fragment molecular orbital (Q-EFMO) method, a hybrid framework that evaluates selected molecular-fragment correlation energies with the variational quantum eigensolver and assembles them through the size-consistent EFMO energy expression. A Hartree--Fock EFMO calculation supplies monomer references, many-body polarization, and distant-pair EFP interactions; independent VQE/UCCSD calculations provide correlation increments for monomers and selected near-field dimers. Consequently, the maximum quantum register is determined by the largest active fragment or fragment pair rather than by total system size. For a three-layer LiH benchmark in STO-3G, a full 2.5--3.5 \AA{} separation scan shows that two orbital-reduction schemes approach the full CCSD reference monotonically. The more aggressive scheme remains below 1 kcal mol$^{-1}$ for separations of 2.6 \AA{} and greater and reaches 0.10 kcal mol$^{-1}$ at 3.5 \AA{}, using 6 qubits for monomers and 12 for the largest dimers, compared with 14 qubits for the frozen-core dimer baseline. An implementation-specific cost proxy decreases from 1000 to 40, a factor of 25. These noise-free results establish the equations, software path, and resource scaling; broader chemical validation, larger bases, and hardware tests remain necessary.
\end{abstract}

\section{Introduction}
Quantum computers offer a natural representation of correlated electronic wave functions, but practical calculations remain constrained by noise, circuit depth, and the number of reliably controlled qubits \cite{preskill2018,mcardle2020,cao2019}. Fragmentation replaces one large electronic-structure problem with smaller, independently solvable subproblems whose energies are reassembled classically. The fragment molecular orbital (FMO) method organizes monomer, dimer, and higher-body terms in a systematic expansion \cite{gordon2012,fmo_book,kitaura1999}. The effective fragment molecular orbital (EFMO) method combines this real-space decomposition with effective fragment potential (EFP) terms for distant pairs and a self-consistent many-body polarization field \cite{steinmann2010,pruitt2013,pruitt2014,jensen1996,gordon2007,gordon2001}.

Previous quantum Monte Carlo/EFMO studies showed that high-level fragment correlation can be combined with EFMO for intermolecular systems and for fragmentation across covalent bonds \cite{zahariev2019,zahariev2021}. The architecture is equally well suited to quantum computing: monomer and near-field dimer correlation increments can be evaluated on a quantum backend, while the Hartree--Fock EFMO reference, long-range EFP interactions, polarization, and final energy assembly remain classical.

We denote this workflow Q-EFMO. Its central resource property is that the largest monomer or selected dimer active space, rather than the total number of fragments, sets the register size. Increasing system size increases the number of independent quantum jobs, which can be distributed or executed sequentially, but need not increase the qubits required by any one job. The present study defines the working equations and software path and evaluates a three-layer LiH benchmark over eleven interlayer separations. All quantum results are statevector UCCSD/STO-3G calculations and should be interpreted as an internal workflow validation.

Q-EFMO is one layer of a broader resource-reduction hierarchy. Q-EFP can embed an entire fragmented cluster in a first-principles classical environment, while virtual-orbital fragmentation (Q-FVO) can reduce the active virtual space inside each Q-EFMO monomer or dimer. Environment embedding, real-space fragmentation, and orbital-space fragmentation therefore address different contributions to quantum-resource growth.

\section{Theory}
\subsection{FMO and EFMO energy expressions}
For $N$ molecular fragments, the conventional FMO2 energy is
\begin{equation}
E_{\mathrm{FMO2}}=\sum_I E_I+\sum_{I<J}\left(E_{IJ}-E_I-E_J\right),
\label{eq:fmo2}
\end{equation}
where the monomer and dimer energies are evaluated in the appropriate embedding fields. EFMO reorganizes this expression so that close pairs are evaluated quantum mechanically, distant pairs are represented by EFP, and the many-body polarization energy is computed once for the full fragment set. A convenient two-body expression is
\begin{equation}
E_{\mathrm{EFMO2}}=
\sum_I E_I^{0}
+\sum_{\substack{I<J\\R_{IJ}\le R_{\mathrm{cut}}}}\Delta E_{IJ}^{\mathrm{QM}}
+\sum_{\substack{I<J\\R_{IJ}>R_{\mathrm{cut}}}}E_{IJ}^{\mathrm{EFP,np}}
+E_{\mathrm{pol}}^{\mathrm{tot}},
\label{eq:efmo2}
\end{equation}
where $E_I^{0}$ is the isolated-fragment monomer energy used by the EFMO assembly, $E_{\mathrm{pol}}^{\mathrm{tot}}$ is the self-consistent many-body polarization energy, and the nonpolarization distant-pair interaction is
\begin{equation}
E_{IJ}^{\mathrm{EFP,np}}=E_{IJ}^{\mathrm{Coul}}+E_{IJ}^{\mathrm{disp}}+E_{IJ}^{\mathrm{exch-rep}}+E_{IJ}^{\mathrm{CT}}.
\label{eq:efpnp}
\end{equation}
The near-field correction is
\begin{equation}
\Delta E_{IJ}^{\mathrm{QM}}=E_{IJ}^{\mathrm{QM}}-E_I^{0}-E_J^{0}-E_{IJ}^{\mathrm{pol}},
\label{eq:near}
\end{equation}
so that the pair polarization already contained in $E_{\mathrm{pol}}^{\mathrm{tot}}$ is not counted twice. When noninteracting groups are separated to infinite distance, all cross-group pair corrections vanish and Eq.~\ref{eq:efmo2} becomes additive, giving the required size-consistent dissociation limit.

Near-field pairs are selected with a closest-contact distance normalized by van der Waals radii,
\begin{equation}
R_{IJ}=\min_{i\in I,\,j\in J}\frac{|\mathbf r_i-\mathbf r_j|}{R_i^{\mathrm{vdW}}+R_j^{\mathrm{vdW}}},
\label{eq:rij}
\end{equation}
with $R_{\mathrm{cut}}$ chosen for the target chemical regime. The precise cutoff and the selected pair list are reported with the corresponding inputs.

\subsection{Quantum correlation increments}
Q-EFMO uses HF-EFMO as the classical base and replaces or augments selected correlation contributions with quantum results. For a monomer or dimer subproblem $X$, define
\begin{equation}
\Delta E_{\mathrm{corr}}^{Q}(X)=E_{\mathrm{VQE}}(X)-E_{\mathrm{HF}}(X).
\label{eq:corr}
\end{equation}
The two-body Q-EFMO energy can then be written
\begin{equation}
E_{\mathrm{Q\mbox{-}EFMO}}^{(2)}=E_{\mathrm{HF\mbox{-}EFMO}}
+\sum_I\Delta E_{\mathrm{corr}}^{Q}(I)
+\sum_{\substack{I<J\\R_{IJ}\le R_{\mathrm{cut}}}}
\left[\Delta E_{\mathrm{corr}}^{Q}(IJ)-\Delta E_{\mathrm{corr}}^{Q}(I)-\Delta E_{\mathrm{corr}}^{Q}(J)\right].
\label{eq:qefmo}
\end{equation}
The fragment correlation calculations use VQE with the UCCSD ansatz \cite{peruzzo2014,mcclean2016}. Because every monomer and selected-dimer expectation-value problem is independent, the quantum stage is embarrassingly parallel.

\subsection{Orbital reduction and scaling}
Scheme 1 applies a frozen-core approximation. Scheme 2 retains the same chemically important occupied space and additionally removes selected high-lying or weakly coupled virtual orbitals. For the LiH benchmark, monomers require 6 qubits in both schemes, while the largest dimers require 14 qubits in Scheme 1 and 12 in Scheme 2. The truncation reduces the number of UCCSD excitation operators as well as the register size. Because these choices are implementation-specific, the exact orbital lists, selection criteria, and cost model are documented with the corresponding input files.

For a sparse near-field graph, the number of quantum jobs scales approximately as $O(N_{\mathrm{frag}})+O(N_{\mathrm{near}})$. In three-dimensional systems with a fixed local cutoff, $N_{\mathrm{near}}$ can grow approximately linearly with fragment count. The maximum qubit count remains that of the largest active monomer or near-field dimer.

\section{Implementation and computational details}
\begin{figure}[H]
\centering
\includegraphics[width=0.98\textwidth]{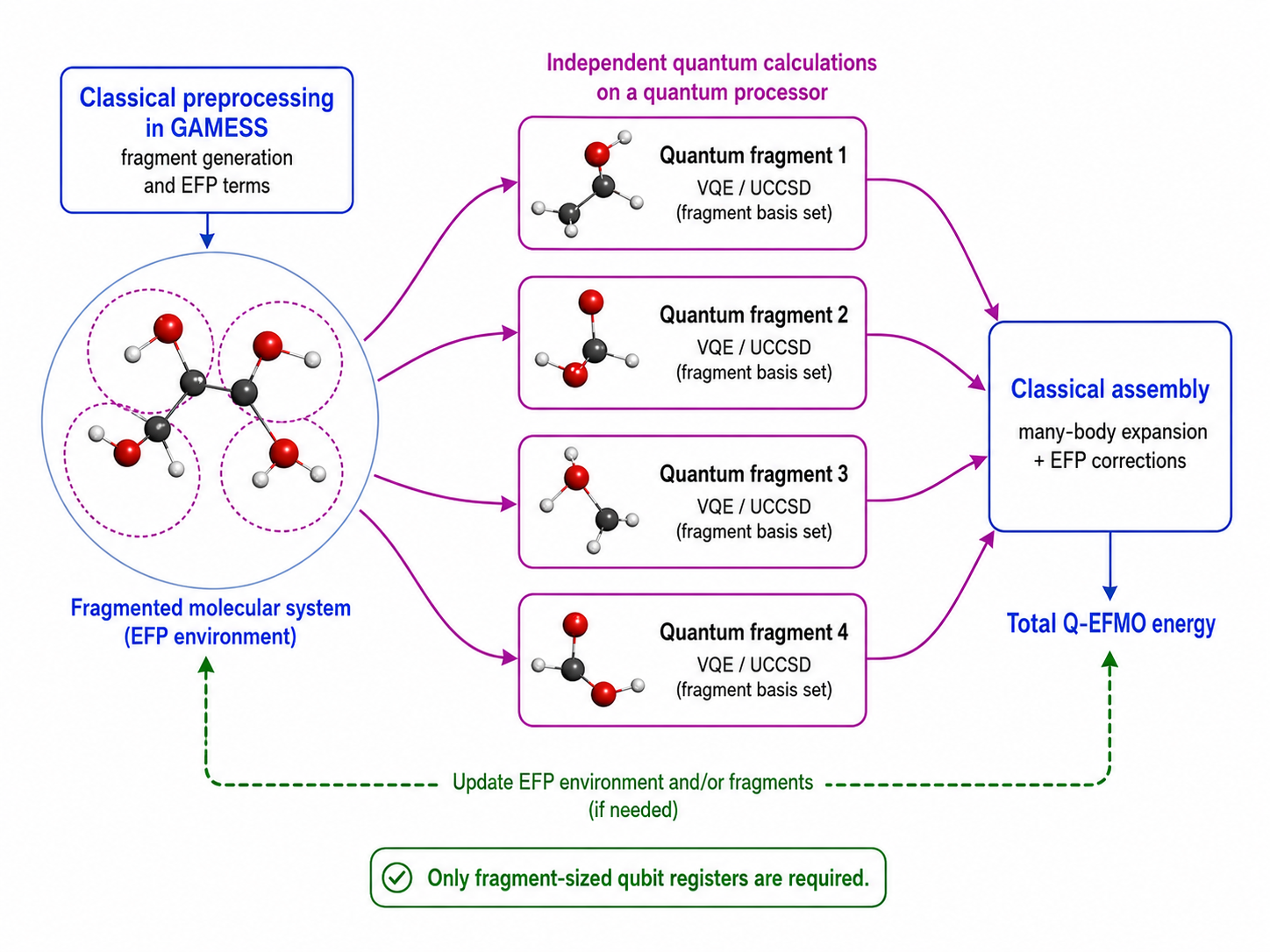}
\caption{Q-EFMO workflow. GAMESS partitions the molecular system and supplies the EFP/EFMO classical information. Independent fragment-sized VQE/UCCSD calculations are performed for selected monomers or dimers, and their correlation increments are combined through the EFMO many-body expression. The feedback path denotes optional updates to the fragment or EFP environment; only fragment-sized quantum registers are required.}
\label{fig:qefmo_workflow}
\end{figure}

The Q-EFMO implementation follows six stages (Figure~\ref{fig:qefmo_workflow}). (1) GAMESS performs a whole-system HF-EFMO calculation and generates fragment, geometry, polarization, and pair information. (2) Python utilities parse the output and select monomers and near-field dimers. (3) Independent fragment HF calculations generate molecular orbitals and integrals. (4) Q-GAMESS converts those integrals to quantum-backend inputs. (5) VQE/UCCSD or another correlated solver evaluates the selected fragment increments. (6) The results are collected and assembled with Eq.~\ref{eq:qefmo}. Fragment calculations in stages 3 and 5 are independent and can be scheduled with workflow tools such as FireWorks or Balsam \cite{gamess1993,gordon2005,qiskit2021,jain2015,salim2019}.

\begin{figure}[H]
\centering
\includegraphics[width=0.47\textwidth]{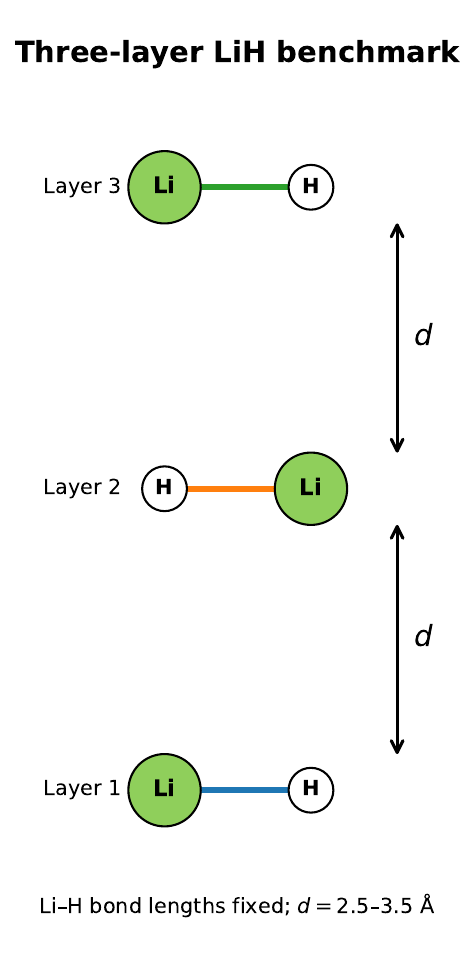}
\caption{Three-layer LiH benchmark. The Li--H bond lengths are fixed, and the separation $d$ between adjacent layers is scanned from 2.5 to 3.5 \AA.}
\label{fig:lih_stack}
\end{figure}

The validation system is a stack of three LiH units (Figure~\ref{fig:lih_stack}). Intramolecular Li--H bond lengths were held at their MP2/6-311++G(3df,3p) optimized values, while the interlayer separation $d$ was varied from 2.5 to 3.5 \AA{} in 0.1 \AA{} increments. Fragment calculations and the full classical reference used STO-3G. Quantum calculations were noise-free statevector UCCSD simulations. Errors are reported relative to the full-system CCSD energy at the same geometry and basis. Total energies in Table~\ref{tab:qefmo_energy} are reported in hartree, and absolute errors are reported in kcal mol$^{-1}$ ($1$ hartree $=627.509474$ kcal mol$^{-1}$).

\section{Results and discussion}
\subsection{Full interlayer-separation scan}
\begin{figure}[H]
\centering
\includegraphics[width=0.78\textwidth]{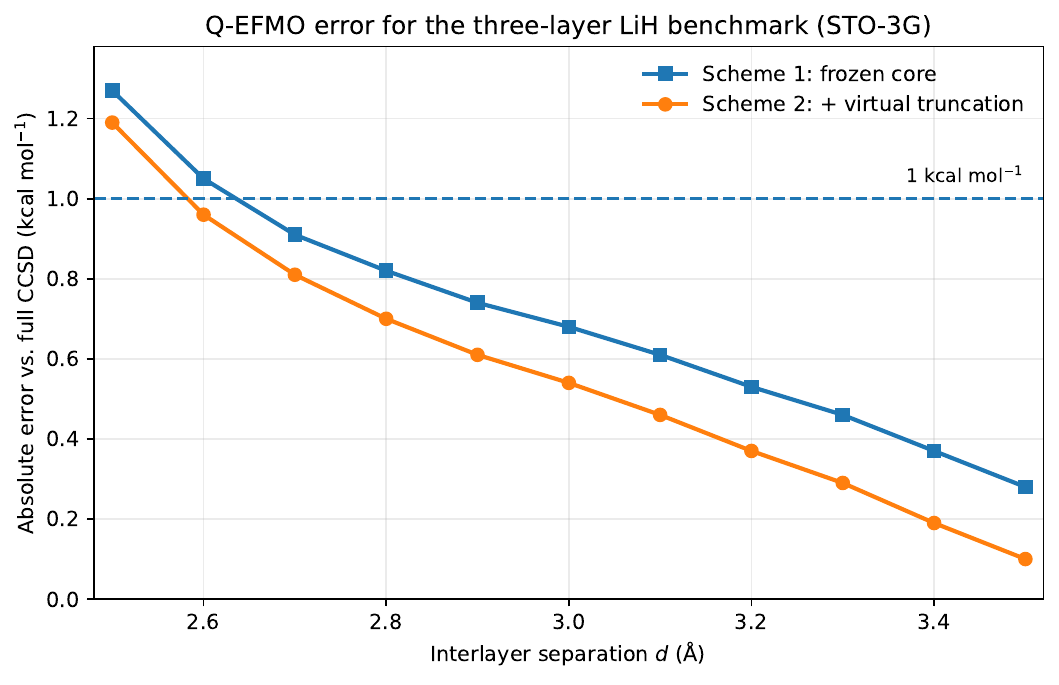}
\caption{Absolute Q-EFMO error relative to the full-system CCSD reference across the complete interlayer scan. Both schemes improve as the layers separate; Scheme 2 remains below 1 kcal mol$^{-1}$ from 2.6 to 3.5 \AA.}
\label{fig:qefmo_error}
\end{figure}

Both orbital-reduction schemes approach the full CCSD reference monotonically as the layers are separated (Table~\ref{tab:qefmo_energy} and Figure~\ref{fig:qefmo_error}). Scheme 2 is systematically closer to the reference throughout the scan. It is below 1 kcal mol$^{-1}$ for all separations at and above 2.6 \AA{}, decreases from 1.19 kcal mol$^{-1}$ at 2.5 \AA{} to 0.10 kcal mol$^{-1}$ at 3.5 \AA{}, and therefore captures the expected weakening of interfragment correlation with separation. Scheme 1 follows the same trend, decreasing from 1.27 to 0.28 kcal mol$^{-1}$.

\begin{table}[H]
\centering
\footnotesize
\caption{Q-EFMO and full CCSD energies for the three-layer LiH stack. Total energies are in hartree; errors are absolute values in kcal mol$^{-1}$.}
\label{tab:qefmo_energy}
\begin{tabularx}{\textwidth}{@{}CCCCCC@{}}
\toprule
$d$ (\AA) & Scheme 1 & Error 1 & Scheme 2 & Error 2 & Full CCSD \\
\midrule
2.5 & $-23.703228$ & 1.27 & $-23.703100$ & 1.19 & $-23.701204$ \\
2.6 & $-23.695818$ & 1.05 & $-23.695674$ & 0.96 & $-23.694144$ \\
2.7 & $-23.689348$ & 0.91 & $-23.689188$ & 0.81 & $-23.687897$ \\
2.8 & $-23.683754$ & 0.82 & $-23.683563$ & 0.70 & $-23.682447$ \\
2.9 & $-23.678957$ & 0.74 & $-23.678750$ & 0.61 & $-23.677778$ \\
3.0 & $-23.674878$ & 0.68 & $-23.674655$ & 0.54 & $-23.673794$ \\
3.1 & $-23.671420$ & 0.61 & $-23.671180$ & 0.46 & $-23.670447$ \\
3.2 & $-23.668439$ & 0.53 & $-23.668184$ & 0.37 & $-23.667595$ \\
3.3 & $-23.665906$ & 0.46 & $-23.665635$ & 0.29 & $-23.665173$ \\
3.4 & $-23.663722$ & 0.37 & $-23.663436$ & 0.19 & $-23.663133$ \\
3.5 & $-23.661826$ & 0.28 & $-23.661539$ & 0.10 & $-23.661380$ \\
\bottomrule
\end{tabularx}
\vspace{0.35em}\parbox{0.98\textwidth}{\footnotesize Scheme 1 uses the frozen-core baseline; Scheme 2 adds virtual-space truncation.}
\end{table}

\subsection{Orbital-reduction comparison}
At $d=3.0$ \AA{}, Scheme 2 reduces the largest fragment register from 14 to 12 qubits and lowers the error from 0.68 to 0.54 kcal mol$^{-1}$ (Table~\ref{tab:qefmo_reduction}). The implementation-reported cost proxy decreases from 1000 to 40, a factor of 25. Both schemes use 6-qubit monomers at 0.2 node-hours per monomer, and the dimer estimate decreases from 48 to 2 node-hours. These figures document the present workflow rather than a hardware-independent complexity or wall-time prediction; they refer to the specific machine, fragment count, orbital lists, scheduling assumptions, and proxy definition of this implementation.

\begin{table}[H]
\centering
\small
\caption{Orbital-reduction schemes for the LiH benchmark.}
\label{tab:qefmo_reduction}
\begin{tabularx}{\textwidth}{@{}YCCCC@{}}
\toprule
Scheme & Virtual orbitals (mon./dim.) & Qubits (mon./dim.) & Node-hours (mon./dim.) & Total proxy \\
\midrule
1: frozen core & 4 / 7--8 & 6 / 14 & 0.2 / 48 & 1000 \\
2: + virtual truncation & 4 / 7--9 & 6 / 12 & 0.2 / 2 & 40 \\
\bottomrule
\end{tabularx}
\vspace{0.35em}\parbox{0.98\textwidth}{\footnotesize Node-hour and total-proxy values are implementation-specific estimates for the present workflow.}
\end{table}

\subsection{Hardware considerations and scope}
Transition to quantum hardware requires finite-shot uncertainty, device noise, circuit depth, measurement grouping, and error mitigation to be treated explicitly. Because fragment jobs are independent, mitigation and ansatz choices can be tuned fragment by fragment and distributed across available processors. Adaptive ansatz construction, symmetry verification, orbital optimization, and zero-noise extrapolation are natural next steps.

The current evidence is deliberately narrow: one layered ionic model, one minimal basis, and statevector calculations. STO-3G captures only a limited part of dynamical correlation, and a single geometry family does not establish transferability to hydrogen-bonded clusters, covalent fragmentation, or chemically heterogeneous systems. Broader validation should vary $R_{\mathrm{cut}}$, separate fragmentation error from UCCSD ansatz error, test larger basis sets, and archive raw fragment energies and assembly scripts. Natural-orbital or entanglement-guided active spaces and the Q-FVO hierarchy may further reduce the fragment registers.

\section{Conclusions}
Q-EFMO combines the EFMO energy architecture with quantum evaluation of selected fragment correlation increments. The qubit requirement is determined by the largest active monomer or near-field dimer rather than by total system size, while independent fragment jobs can be executed in parallel. Across the complete 2.5--3.5 \AA{} LiH scan, both orbital schemes converge monotonically toward full CCSD; Scheme 2 reaches 0.10 kcal mol$^{-1}$ error at 3.5 \AA{}, uses 6-qubit monomers and 12-qubit dimers, and reduces the implementation cost proxy by a factor of 25 relative to the 14-qubit frozen-core dimer baseline. These results establish a functioning proof-of-workflow. Combined with Q-EFP embedding and Q-FVO virtual-space reduction, Q-EFMO provides the real-space layer of a hierarchical strategy for larger molecular environments.

\section{Acknowledgments}
This work was supported in part by the U.S. Department of Energy, Office of Science, Basic Energy Sciences. This manuscript has been authored by UT-Battelle, LLC, under Contract No. DE-AC05-00OR22725 with the U.S. Department of Energy. This work was also supported by the U.S. Department of Energy, Office of Science, through Ames National Laboratory under Contract No. DE-AC02-07CH11358. The authors acknowledge partial support by ORNL LDRD and VSO programs. The U.S. Government retains, and the publisher by accepting the article for publication acknowledges, a nonexclusive, paid-up, irrevocable, worldwide license to publish or reproduce the published form of this manuscript, or allow others to do so, for U.S. Government purposes. This research used resources of the Oak Ridge Leadership Computing Facility (Frontier; Director's Discretionary allocation CHM238) and the National Energy Research Scientific Computing Center under award m4621 (ERCAP0036406). Computational resources were also provided by Iowa State University.

\section{Code and data availability}
Q-GAMESS is available at \url{https://github.com/fzahari/Q-GAMESS}. Input geometries, fragment definitions, the EFMO cutoff and pair list, active-orbital lists, raw fragment energies, optimizer and backend settings, and the cost-proxy definition are available from the corresponding author upon reasonable request.

\FloatBarrier

\end{document}